\documentclass{article}
\usepackage{frascatiphys,here,graphicx,subfigure}
\begin{document}
\title{SYSTEMATIC ERRORS OF BOUND-STATE PARAMETERS
EXTRACTED BY MEANS OF SVZ SUM RULES}
\author{Wolfgang Lucha
\\{\em Institute for High Energy Physics,
Austrian Academy of Sciences,}
\\{\em Nikolsdorfergasse 18, A-1050,
Vienna, Austria}
\\Dmitri Melikhov
\\{\em Institute for High Energy
Physics, Austrian Academy of Sciences,}
\\{\em Nikolsdorfergasse
18, A-1050, Vienna, Austria}\\
{\em and}\\{\em Nuclear Physics
Institute, Moscow State University,}
\\{\em 119991, Moscow, Russia}\\
Silvano Simula\\
{\em INFN, Sezione di Roma III,}
\\{\em
Via della Vasca Navale 84, I-00146, Roma, Italy}}
\maketitle
\baselineskip=11.6pt

\begin{abstract}
This talk presents the results of our study of
systematic errors of the ground-state parameters obtained by
Shifman--Vainshtein--Zakharov (SVZ) sum rules. We use the
harmonic-oscillator potential model as an example: in this case we
know the exact solution for the polarization operator, which
allows us to obtain both the OPE to any order and the parameters
(masses and decay~constants)~of the bound states. We extract the
parameters of the ground state by making use of the standard
procedures of the method of QCD sum rules, and compare the
obtained results with their known exact values. We show that if
the continuum contribution to the polarization operator is not
known and is modelled by some effective continuum threshold, the
standard procedures adopted in sum rules do not allow one to gain
control over the systematic errors of the extracted ground-state
parameters.
\end{abstract}
\baselineskip=14pt

\newpage
A QCD sum-rule calculation of hadron parameters\cite{svz} involves
two steps: one first constructs the operator product expansion
(OPE) series for a relevant correlator and then extracts the
parameters of the ground state by a numerical procedure. Each of
these steps leads to certain uncertainties in the final result.

The first step lies entirely within QCD and, in the case of SVZ
sum rules, allows for a rigorous treatment of the uncertainties:
the correlator is not known precisely because of uncertainties in
quark masses, condensates, $\alpha_s$, etc., but~all corresponding
errors in the correlator may be controlled. [Complications arising
in light-cone sum rules are discussed in our second
talk\cite{talk2}.]

The second step lies beyond QCD: even if several terms of the OPE
for the correlator were known precisely, the hadronic parameters
might be extracted~by a sum rule only within some error, which may
be treated as a systematic error~of the method.

Here we present the results of our recent study of systematic
uncertainties of the sum-rule procedures\cite{lms_sr,lms_lcsr}. To
this end, a quantum-mechanical harmonic-oscillator (HO) potential
model is a perfect tool: in this case both the spectrum of bound
states (i.e., masses and wave functions) and the exact correlator
(and hence its OPE to any order) are known precisely. Therefore,
one may apply the sum-rule machinery for extracting parameters of
the ground state and test the accuracy of the extracted values by
comparing with the known exact results. In this way the accuracy
of the method can be probed. For a detailed discussion~of various
aspects of sum rules in quantum mechanics, we refer to
Refs.~[5--9].

To illustrate the essential features of the QCD calculation, we
consider a non-relativistic model with a confining potential,
\begin{eqnarray}
\label{1.1} V(r)=\frac{m\omega^2\mathbf{r}^2}{2},\qquad
r=|\mathbf{r}|,
\end{eqnarray}
and analyze the Borel transform $\Pi(\mu)$ of the polarization
operator $\Pi(E),$~which gives the evolution operator in the
imaginary time $1/\mu$:
\begin{eqnarray}
\Pi(\mu)=\left(\frac{2\pi}{m}\right)^{3/2}\left\langle
\mathbf{r}_f=\mathbf{0}\left|\exp\left(-\frac{H}{\mu}\right)\right|
\mathbf{r}_i=\mathbf{0}\right\rangle.
\end{eqnarray}
For the HO potential (\ref{1.1}), the exact analytic expression
for $\Pi(\mu)$ is well known:
\begin{eqnarray}
\label{piexact}
\Pi(\mu)=\left(\frac{\omega}{\sinh(\omega/\mu)}\right)^{3/2}.
\end{eqnarray}
Expanding the above expression in inverse powers of $\mu$, we get
the OPE series
\begin{eqnarray}
\label{piope} \Pi_{\rm OPE}(\mu)&\equiv&
\Pi_{0}(\mu)+\Pi_{1}(\mu)+\Pi_{2}(\mu)+\cdots\nonumber\\[1ex]
&=&\mu^{3/2}
\left(1-\frac{\omega^2}{4\mu^2}+\frac{19}{480}\frac{\omega^4}{\mu^4}
+\cdots\right);
\end{eqnarray}
higher power corrections may be derived from the exact result (\ref{piexact}).

The ``phenomenological'' representation for $\Pi(\mu)$ is obtained
by using the basis of hadronic eigenstates of the model, namely,
\begin{eqnarray}
\label{piphen1} \Pi(\mu)=\sum_{n=0}^\infty R_n
\exp\left(-\frac{E_n}{\mu}\right),
\end{eqnarray}
where $E_n$ is the energy of the $n$th bound state and 
$R_n$ [the square of the leptonic decay constant of the $n$th bound state] 
is given by
\begin{eqnarray}
R_n=\left(\frac{2\pi}{m}\right)^{3/2}|\Psi_n(\mathbf{r}=\mathbf{0})|^2.
\end{eqnarray}
For the lowest states, one finds from (\ref{piexact})
\begin{eqnarray}
\label{E0} E_0=\frac{3}{2}\omega,\ R_0=2\sqrt{2}\omega^{3/2},
\qquad E_1=\frac{7}{2}\omega,\
R_1=3\sqrt{2}\omega^{3/2},\qquad\ldots.
\end{eqnarray}
The sum rule is just the equality of the correlator calculated in
the ``quark''~basis and in the ``hadron'' basis:
\begin{eqnarray}
&&R_0\exp\left(-\frac{E_0}{\mu}\right)+\int\limits_{z_{\rm
cont}}^\infty{\rm d}z\,\rho_{\rm
phen}(z)\exp\left(-\frac{z}{\mu}\right)\nonumber\\
[1ex]
&&\hspace{1cm}=\int\limits_{0}^{\infty}{\rm
d}z\,\rho_0(z)\exp\left(-\frac{z}{\mu}\right)+\mu^{3/2}
\left(-\frac{\omega^2}{4\mu^2}
+\frac{19}{480}\frac{\omega^4}{\mu^4} +\cdots\right).\label{sr}
\end{eqnarray}
Following Ref.~[1], we use explicit expressions for the power
corrections, but for the zeroth-order free-particle term we use
its expression in terms of the spectral integral.

Let us introduce an ``effective'' continuum threshold $z_{\rm
eff}(\mu)$, different from the physical $\mu$-independent
continuum threshold $z_{\rm cont}$, by the relation
\begin{eqnarray}
\label{zeff} 
\Pi_{\rm cont}(\mu)=\int\limits_{z_{\rm
cont}}^\infty {\rm d}z\,\rho_{\rm phen}(z)\exp\left(-\frac{z}{\mu}\right)
=\int\limits_{z_{\rm eff}(\mu)}^\infty{\rm d}z\,\rho_{0}(z)\exp\left(-\frac{z}{\mu}\right).
\end{eqnarray}
The spectral densities $\rho_{\rm phen}(z)$ and $\rho_{0}(z)$ are
different functions. Thus the two sides of (\ref{zeff}) can be
equal to each other only if the effective continuum threshold,
$z_{\rm eff}(\mu)$, depends on $\mu$ in an appropriate way. In
our model, we can calculate $\Pi_{\rm cont}$ precisely and,
therefore, we can obtain the function $z_{\rm eff}(\mu)$ by
solving (\ref{zeff}). In the general case of an actual QCD
sum-rule analysis, the effective continuum threshold is not known
and constitutes one of the essential fitting parameters.

Making use of (\ref{zeff}), we now rewrite the sum rule (\ref{sr})
in the form
\begin{eqnarray}
\label{sr2} R_0\exp\left(-\frac{E_0}{\mu}\right)=\Pi(\mu,z_{\rm
eff}(\mu)),
\end{eqnarray}
where the cut correlator $\Pi(\mu,z_{\rm eff}(\mu))$ reads
\begin{eqnarray}
\label{cut} 
&&\Pi(\mu,z_{\rm eff}(\mu))
\nonumber\\
&&\hspace{.4cm}
\equiv
\frac{2}{\sqrt{\pi}}\int\limits_{0}^{z_{\rm eff}(\mu)}{\rm d}z\,
\sqrt{z}\exp\left(-\frac{z}{\mu}\right)+\mu^{3/2}
\left(-\frac{\omega^2}{4\mu^2}
+\frac{19}{480}\frac{\omega^4}{\mu^4}+\cdots\right).
\end{eqnarray}
As is obvious from (\ref{sr2}), the cut correlator satisfies the
equation
\begin{eqnarray}
\label{e0a} -\frac{{\rm d}\log \Pi(\mu,z_{\rm eff}(\mu))}{{\rm d}(1/\mu)}=E_0.
\end{eqnarray}
The cut correlator $\Pi(\mu,z_{\rm eff}(\mu))$ is the
quantity that actually governs the extraction of the ground-state
parameters.

The ``fiducial''\cite{svz} range of $\mu$ is defined as the range where, on
the one hand, the~OPE reproduces the exact expression with better than
some chosen accuracy (for instance, within, say, 0.5\%) and, on
the other hand, the ground state is expected to give a sizable
contribution to the correlator. If we include only the first three
power corrections (that is, $\Pi_1$, $\Pi_2$, and $\Pi_3$) we must
require $\omega/\mu<1.2$. Since we know the ground-state
parameters, we fix $\omega/\mu>0.7$, where the ground state contributes 
more than 60\% of the full correlator. So our fiducial range
is $0.7<\omega/\mu<1.2$.

We shall be interested in situations where the hadronic continuum
is {\em not\/} known --- which is typical for heavy-hadron physics
and in the discussion of~the properties of exotic hadrons. Can we 
extract the ground-state parameters? 

We denote the values of the
ground-state parameters extracted from the sum rule (\ref{sr2}) by 
$E$ and $R$. The notations $E_0$ and $R_0$ are reserved for the exact values. 
\begin{figure}[t]
\begin{center}
\includegraphics[width=6.5cm]{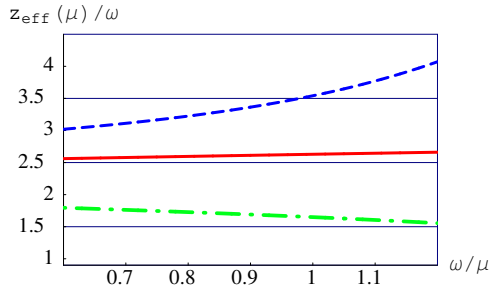}
\caption{\label{Fig:2}The effective continuum threshold $z_{\rm
eff}(\mu)$ obtained by solving (\ref{sr}) for $E=E_0$ and $R=0.7
R_0$ [long-dashed (blue) line], $R=R_0$ [solid (red) line] and
$R=1.15 R_0$ [dash-dotted (green) line]. }
\end{center}
\end{figure}
In many interesting cases the ground-state energy may be
determined, e.g., from experiment. However, setting $E=E_0$ does
not help: still, for any~$R$ within a broad range, one finds 
a function $z_{\rm eff}(\mu,R)$ [Fig.~\ref{Fig:2}] which solves
the~sum rule (\ref{sr2}) {\it exactly}. Therefore, we conclude
that {\it in a limited range of $\mu$ the OPE alone cannot say
much about the ground-state parameters}. What really matters is
the continuum contribution, or, equivalently, $z_{\rm eff}(\mu)$.
Only by making some assumptions about $z_{\rm eff}(\mu)$ one is able to extract $R$. 

Typically, one assumes
$z_{\rm eff}(\mu)$ to be constant and imposes some criteria to fix
its value. Rigorously speaking, a {\em constant\/} effective
continuum threshold $z_{\rm eff}(\mu)=z_c={\rm const}$ is
incompatible with the sum rule (\ref{sr2}). Nevertheless, such an
Ansatz may work well, especially in our HO model: As seen from~
Fig.~\ref{Fig:2}, the exact $z_{\rm eff}(\mu)$ is almost flat in
the fiducial interval. Therefore, the HO model represents a very
favorable situation for applying the QCD sum-rule machinery.

\begin{figure}[ht]
\begin{center}
\begin{tabular}{cc}
\includegraphics[width=6.7cm]{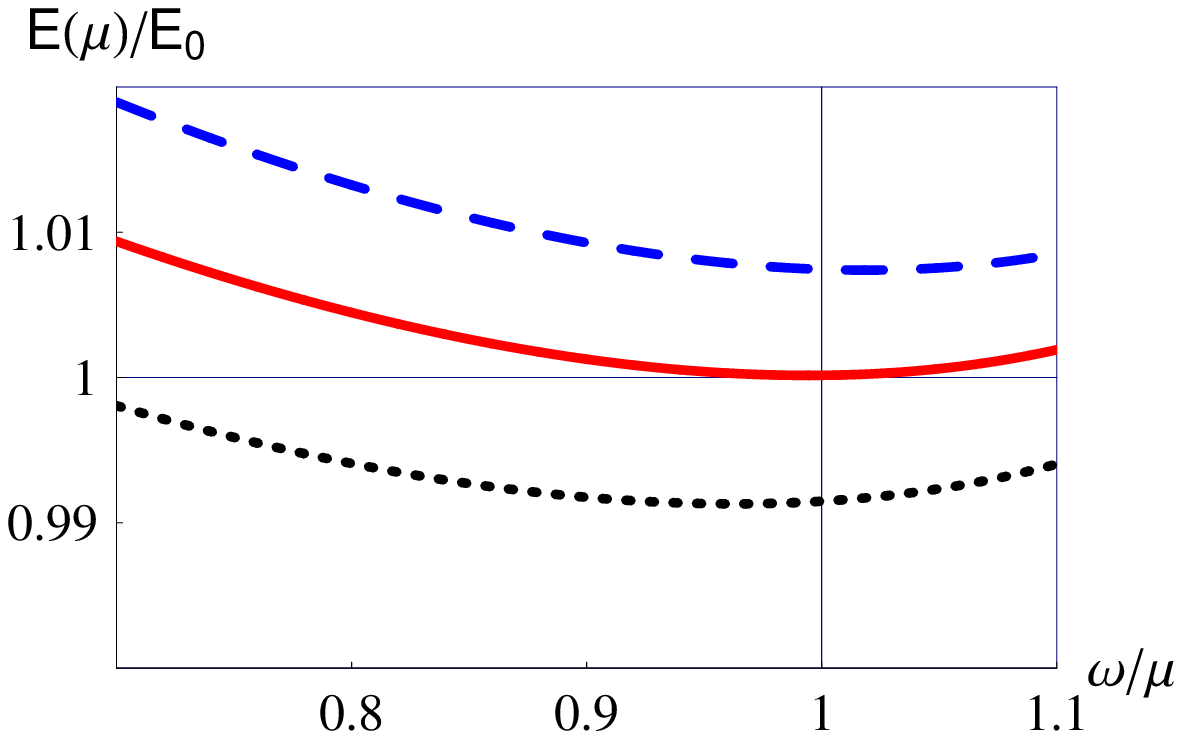}\\(a)\\
\includegraphics[width=6.7cm]{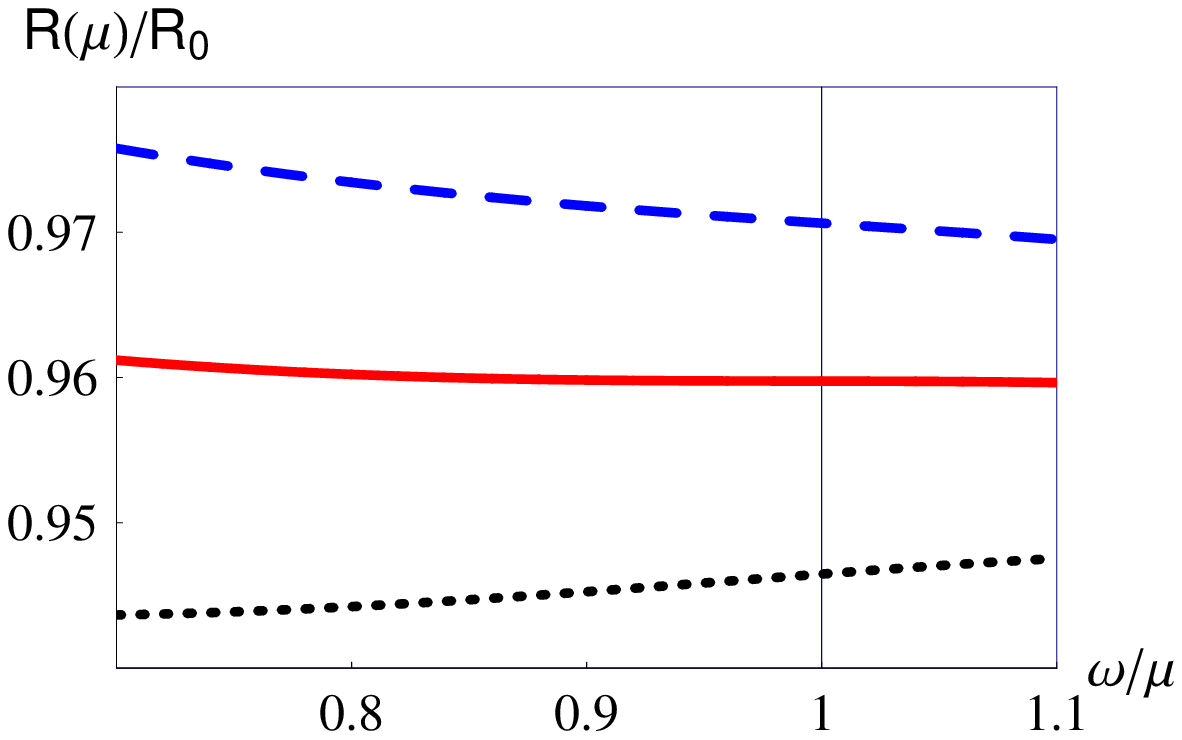}\\(b)
\end{tabular}
\end{center}
\caption{\label{Fig:3}Constant effective
continuum threshold $z_c$: $E(\mu)$ for three different values of
$z_c$ (a) and the corresponding $R(\mu)$ (b).}
\end{figure}
Now, how to determine $z_c$? A widely used procedure\cite{jamin} is to calculate
\begin{eqnarray}
E(\mu,z_c)\equiv -\frac{{\rm d}\log\Pi(\mu,z_c)}{{\rm d}(1/\mu)},
\end{eqnarray}
which now depends on $\mu$ due to approximating $z_{\rm eff}(\mu)$
by a constant. Then,~one determines $\mu_0$ and $z_c$ as the
solution to the system of equations
\begin{eqnarray}
\label{add} E(\mu_0,z_c)=E_0,\qquad
\left.\frac{\partial}{\partial\mu}E(\mu,z_c)\right|_{\mu=\mu_0}=0,
\end{eqnarray}
yielding $z_c=2.454\,\omega$, $\mu_0/\omega=1$ [Fig.~\ref{Fig:3}].
Finally, one takes the value $R(\mu_0,z_c)$ as the sum-rule
estimate for the quantity $R$. The error of $R$ is usually
obtained by looking at the range covered by $R(\mu,z_c)$ when one
allows for a variation of~$\mu$ within the fiducial range.
Following this procedure, one obtains in our~case a good
central-value estimate: $R/R_0=0.96$. Since $R(\mu,z_c)$ is
extremely stable in the fiducial range, one expects its true value
to be rather close to the extracted value and, accordingly,
assigns a very small error to the sum-rule estimate.

Note, however, a dangerous point: (i) a perfect description of
$\Pi(\mu)$ with~an accuracy better than 1\%, (ii) a deviation of
$E(\mu,z_c)$ from $E_0$ at the level of~only 1\%, and (iii) an
extreme stability of $R(\mu)$ in the entire fiducial range
conspire~to lead to a 4\% error in the extracted value of $R$!
Clearly, this error could not be guessed on the basis of the other
numbers obtained, and it would be wrong to try to estimate the
error from, e.g.,~the~range covered by $R$ when varying the Borel
parameter $\mu$ within the fiducial~interval.

\newpage 
Let us summarize the lessons we have learnt from the above
investigation:

\vspace{.13cm}
\noindent 1. The knowledge of the correlator to any
accuracy within a limited range of~the Borel parameter $\mu$ is
not sufficient for an extraction of the ground-state
parameters since rather different models for the correlator,
generically of the form~of a ground state plus an effective
continuum, lead to the same correlator.

\vspace{.13cm}
\noindent 2. Modelling the hadron continuum by a
{\it constant\/} effective continuum threshold $z_c$ allows one to
determine the value of $z_c$ by, e.g., requiring the average
energy $E(\mu)$ to be close to $E_0$ in the region of stability 
of the sum rule.
In the model under discussion this leads to a good estimate,
$R/R_0=0.96$, with almost $\mu$-independent $R$.~The unpleasant
feature of this extraction procedure is that the deviation of $R$
from $R_0$ is much larger than the variations of $E(\mu)$ and
$R(\mu)$ over the fiducial interval of $\mu$. In particular, it
would be wrong to assign the systematic error on the~basis of the
range covered by $R(\mu)$ when $\mu$ is varied within the fiducial
interval. This means that the standard procedures adopted in QCD
sum rules do not allow one to control the systematic errors.
Consequently, no rigorous systematic errors for hadronic 
parameters extracted by sum rules can be provided. Let us also
stress that the independence of the extracted values of the hadron
parameters from the Borel mass $\mu$ does not guarantee the
extraction of their true values.

\vspace{.13cm}
Finally, in the model under consideration sum rules
provide a rather good estimate for $R_0$, even though its error
cannot be determined on the basis of the standard procedures
adopted in sum-rule analyses. This may be a consequence of the
following features of the model: (i) a large gap between ground
state~and the first excitation contributing to the sum rule; (ii)
an almost constant exact effective continuum threshold. Whether or
not the same good accuracy may~be achieved in QCD, where the
features mentioned above are absent, is not obvious at all and
requires more detailed investigations.

We would like to point out that with respect to the problem of
assigning systematic errors to the extracted hadron parameters,
the method of QCD~sum rules faces very similar problems as the
application of approaches based on the constituent quark picture:
for instance, the relativistic dispersion approach\cite{m} yields
very successful predictions for the form factors of exclusive $D$
decays and provides many predictions for the form factors of weak
decays of $B$ mesons\cite{m1}. However, assigning rigorous errors
to these predictions could not be done so~far.

\newpage\noindent{\bf Acknowledgements.} 
We thank R.~Bertlmann, B.~Grinstein, and B.~Stech~for interesting 
discussions and inspiring comments, and the Austrian Science Fund (FWF) for
financial support under project P17692.

\end{document}